%% file: IF07_topic4.tex
\documentclass[11pt]{article}
\usepackage{graphicx}
\usepackage[margin=1.25in]{geometry}
\usepackage[usenames,dvipsnames]{color}
\usepackage{url}
\usepackage[affil-it]{authblk}
\usepackage[colorlinks = true,
            linkcolor = blue,
            urlcolor  = blue,
            citecolor = blue,
            anchorcolor = blue]{hyperref}

\newenvironment{Abstract}{\begin{quotation} \begin{center}
                       ABSTRACT
     \end{center}\bigskip  }{\end{quotation}}

\input workshopsymbols.tex

\newcommand\snowmass{
    \begin{center}\rule[-0.2in]{\hsize}{0.01in} \\
        \rule{\hsize}{0.01in} \\
        \vskip 0.1in Submitted to the Proceedings of the US Community Study \\ 
        on the Future of Particle Physics (Snowmass 2021) \\ 
        \rule{\hsize}{0.01in} \\
        \rule[+0.2in]{\hsize}{0.01in} 
    \end{center}
}

\title{\bf Electronics for Fast Timing}
\date{March 31, 2022}

\author[1]{D. Braga}
\author[2]{G. Carini}
\author[2]{G. Deptuch}
\author[4]{A. Dragone}
\author[1]{F. Fahim}
\author[5]{K. Flood}
\author[2]{G. Giacomini}
\author[2]{D. Górni}
\author[1]{R. Lipton}
\author[4]{B. Markovic}
\author[3]{S. Mazza}
\author[2]{S. Miryala}
\author[1]{P. Rubinov}
\author[6]{G. Saffier-Ewing}
\author[3]{H. Sadrozinski}
\author[4]{A. Schwartzman}
\author[3]{A. Seiden}
\author[1]{Q. Sun}
\author[1]{T. Zimmerman}

\affil[1]{Fermi National Accelerator Laboratory, Batavia, IL 60510, USA}
\affil[2]{Brookhaven National Laboratory, Upton, 11973, NY, USA}
\affil[3]{SCIPP, University of California Santa Cruz, Santa Cruz, CA 95064, USA}
\affil[4]{SLAC National Accelerator Laboratory; Menlo Park, California 94025, USA}
\affil[5]{Nalu Scientific LLC}
\affil[6]{Anadyne Inc}

\begin{document}

%\snowmass

%{\let\newpage\relax\maketitle}

\maketitle

%\begin{document}
%\maketitle
%\medskip
%\Address{ }
%\medskip

 \begin{Abstract}
\noindent Picosecond-level timing will be an important component of the next generation of particle physics detectors. The ability to add a 4$^{th}$ dimension to our measurements will help address the increasing complexity of events at hadron colliders and provide new tools for precise tracking and calorimetry for all experiments. Detectors are described in detail on other whitepapers. In this note, we address challenges in electronics design for the new generations of fast timing detectors.
\end{Abstract}

\snowmass

\def\thefootnote{\fnsymbol{footnote}}
\setcounter{footnote}{0}

\section{Introduction}
Time resolution in particle physics has steadily advanced from microseconds in the 1950s to nanoseconds in the 60s to picoseconds today. This progression has allowed our detectors to utilize the steady increase in instantaneous luminosity in colliders. This has provided the ability to perform precision measurements, search for rare decays, and complete the Standard Model. Time resolutions shorter than track propagation or shower development times can give us insight into the characteristics of complex physics events beyond our current capabilities. Time is crucial for background rejection in dark matter searches and neutrino detectors.  As resolution continues to increase, time will likely become an equal partner to position measurement in particle tracking and particle flow event fitting. 

All of this has been enabled both by the rapid and continuous advance of fast electronics and the ability to generate fast signals with good signal/noise from a variety of solid state sensors, photodetectors, and micropattern gas-based detectors. A "poster child" for the combination of time resolution, position resolution, and radiation hardness, may be the Muon Collider, where the signal must be isolated from a sea of out-of-time beam-induced-background \cite{https://doi.org/10.48550/arxiv.2203.07224}.

\section{Constraints and Design Tradeoffs}
\subsection{Signal and Noise}
In a system where the timing resolution is dominated by noise-associated jitter, we can express the 
time resolution in terms of detector and amplifier parameters as:
\begin{equation}
\sigma _t \approx \sigma _{n}\frac{\delta V}{\delta t}  \approx t_r \frac{noise}{signal} , \sigma_n^2 \approx \frac{C_L^2 (4kTA)}{g_m T_a}, 
\sigma_t  \approx \frac{C_L}{\sqrt{g_m t_a}} \frac{\sqrt{t_a^2 + t_d^2}}{signal}
\end{equation} \cite{Cartiglia:2016sjr,Spieler:1982jf}
Where $\sigma_t$ is the time resolution, $\sigma_n$ is the system noise, $\frac{\delta V}{\delta t}$ is related to the amplifier and detector rise times, $t_a$ and $t_d$, 
$C_L$ is the load capacitance, $g_m$ is the front end transistor transductance and $4kTA$ is associated with the thermal noise of the amplifier.
 To achieve good time resolution, we want to maximize signal to noise, minimize risetimes, maximize $g_m$ (which is related to front-end transistor 
 current), and minimize capacitance.  
 
 The above considerations assume uniform pulse shapes, simple time walk corrections, and uniform weighting fields.  In a segmented tracking detector delta rays, landau fluctuations, and track angle can all affect the pulse shape and resulting time resolution. A non-uniform weighting field might require analysis of the waveforms in an array of electrodes. In some applications, especially calorimetry, these variations will be large enough to require full waveform digitization to recover time information from a complex signal. This will add to front-end complexity, power and required bandwidth.

\section{Input Signal Sources}
\subsection{Ultrafast Sensor Signals}
Ultra-fast timing depends singularly on the interplay of timing sensor properties and the design of the readout electronics. It is best characterized by the value for the jitter, combining sensor properties like signal rise time $t_{rise}$ and signal height $S$ with properties of the readout electronics like noise N and the bandwidth which will determine the final rise time and signal height depending e.g. on the detector capacitance:
\begin{equation}
\sigma_{jitter} = \frac{N}{dV/dt} \approx \frac{t_{rise}}{S/N}
\end{equation}
(Additional, detector specific contributions to the time resolution exist). 

 \begin{figure}[ht]
\centering
\includegraphics[width=120 mm]{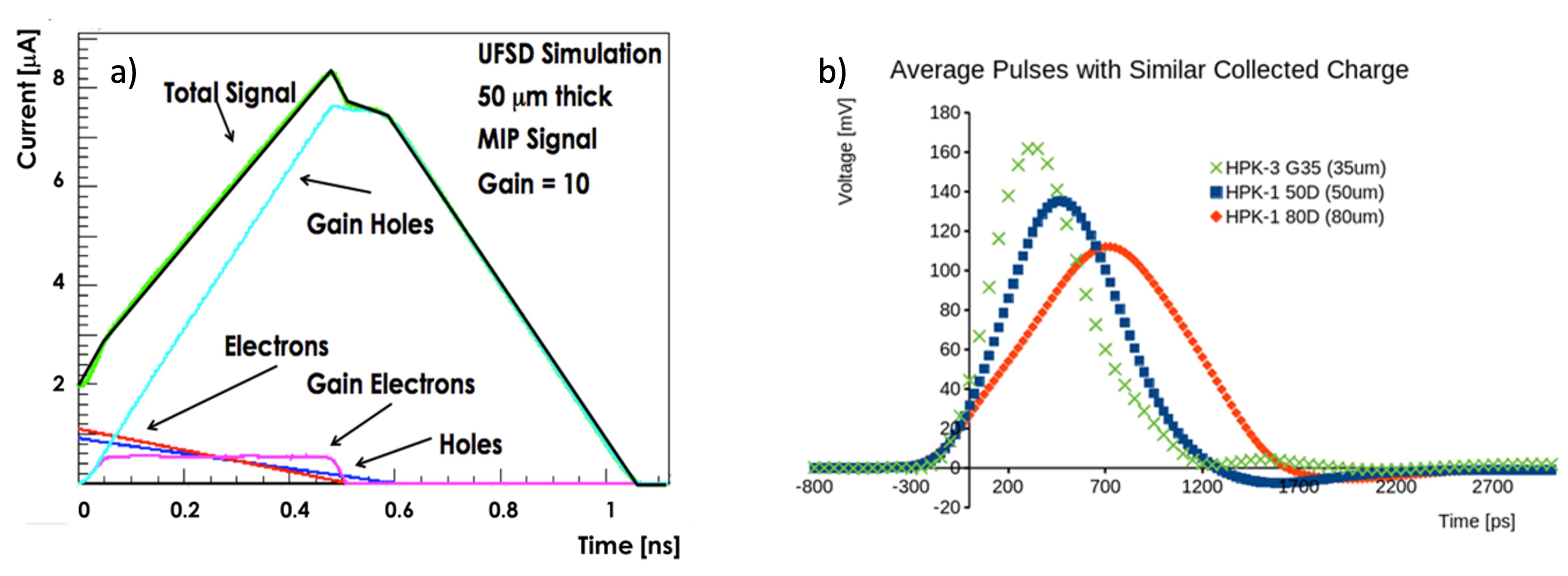}
\caption{Signals in silicon sensors a): composition of pulses of no-gain silicon sensors (sum of “Electrons” and “Holes”) and of LGAD (“Total Signal”); b) LGAD pulse shapes for different sensor thickness }
\label{sig}
\end{figure}

\begin{figure}[ht]
\centering
\includegraphics[width=100 mm]{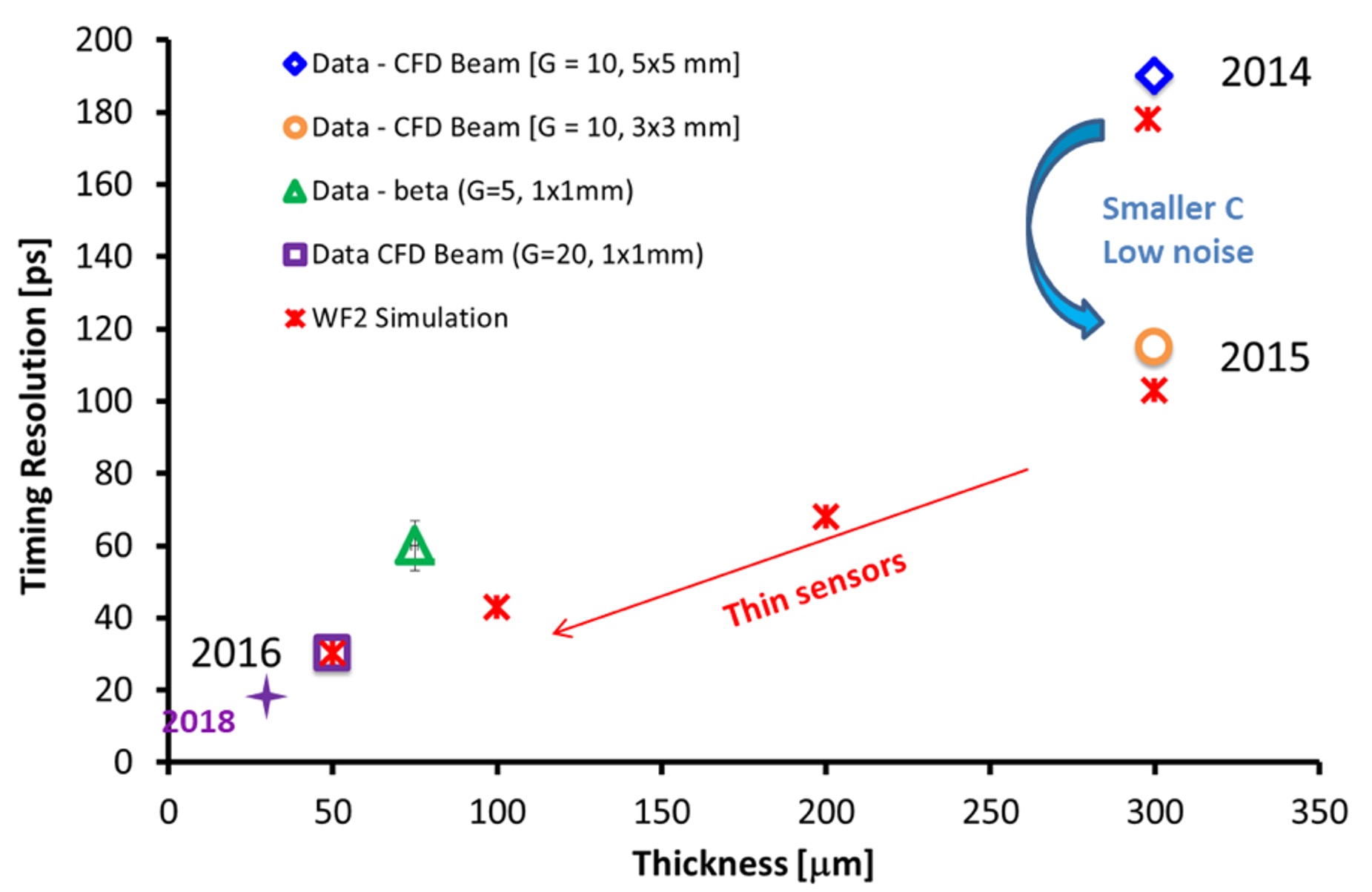}
\caption{Simulated and measured time resolution for LGAD as a function of sensor thickness. }
\label{tim}
\end{figure}

Two options for fast silicon sensors are being developed: sensors with gain, (e.g. Low-gain Avalanche Detectors LGAD) and sensors without gain (e.g. 3D sensors). In order to reduce the jitter, sensors without gain and thus limited signal need to maintain the excellent rise time of the electron collection as shown in Fig. \ref{sig}a, while for LGAD the internal gain boosts the hole signal by a gain of 10-20 by sacrificing part of the rise time of the initial signal production. Both the rise time and the signal height depend on the detector thickness (Fig.\ref{sig}b). For LGAD, the time resolution is limited by the so called “Landau factor” caused by the stochastic variation of the change deposition, which is lower for thinner sensors.Thus the time resolutions depend on the sensor thickness as shown in Fig.\ref{tim}, and favors thin LGAD because of the available gain which can offset the increased rise time\cite{Sadrozinski:2017qpv}.

Both CMS (ETL) and ATLAS (HGTD) are developing large-scale timing layers based on LGAD of 50 um thickness and 1.3 mm pitch. The readout electronics for these sensors are optimized for sensor capacitances of $\sim$ 5 pF and yield $\sim$ 35 ps time resolution at a power of about 5 mW/channel. The next generation of LGAD will require a position resolution of 5-10 um, and a typical pitch will be $< 500 \mu$m, in addition to a factor 2 improved time resolution. This is achieved with AC-LGAD with sparse readout of small pads where charge sharing between pixels allows an interpolation algorithm to reach the position accuracy and a power density similar to the present upgrades under construction. The advantage is a lower pad capacitance of 100 -200 fF. Recent AC-LGAD test beam results show 5$\mu$m and 30 ps resolution\cite{https://doi.org/10.48550/arxiv.2201.07772}. Another LGAD option is currently under investigation: it is the Deep-Junction LGAD where a high-field p/n junction is buried in the substrate a few microns from the patterned electrodes at the interface. Small electrodes lead to lower capacitance, whit timing resolution comparable to LGADs and spatial resolution as found in standard sensors. In general, the internal gain in the sensor permits low-noise, ultra-fast, low-power ASIC design. 
There are ongoing programs for the application of LGAD in approved EIC and PIONEER detectors described in Section 4 of “Four Dimensional Trackers”.

\subsection{Gas Detectors}
Micro-pattern gaseous detectors have the capability to provide excellent time and spatial resolution over large areas. The CERN RD51\cite{RD51} collaboration has explored both detector and electronic systems that have demonstrated $\approx$10 ps resolution. The PICOSEC project\cite{BORTFELDT2018317} uses a Cerenkov radiator and a photocathode to generate the electron signal to achieve 10's of ps resolution. Ps-level system resolution of larger scale MPGD readout requires precise clock distribution, integration of fast amplifiers and TDCs as well as integration of waveform sampling systems. A Scalable Readout System (SRS and SRSe) has been developed which provides both vertical links between front end ASICs and readout and horizontal links among detector systems\cite{6154414}.  The needs for micropattern gas detectors and timing electronics are summarized in the Snowmass Whitepaper on "Micro Pattern Gaseous Detectors for Nuclear Physics"\cite{Barbosa2022Snowmass2I}. R\&D is summarized in the whitepaper "MPGDs: Recent advances and current R\&D" \cite{https://doi.org/10.48550/arxiv.2203.06562}.

\subsection{Pixel Detectors}
Silicon diode-based can achieve picosecond-level timing by careful management 
and understanding of pulse shape and signal/noise. The NA62 Gigatracker achieved 65 ps resolution with 300 $\mu$m pixel size\cite{AglieriRinella:2019eri}.  A pixelated detector can have $>100$ times less load capacitance than the current generation of LGADs, with consequently lower noise. The lower noise can largely make up for loss of signal gain and slower rise time than for LGADs.  Providing sufficient power to the front end to maintain time resolution is a principle challenge in a system with small (i.e. $25\mu$m) pixels.

3D sensors have been proposed as an alternate strategy to achieve fast timing in 
silicon diodes. They can achieve very fast signals, but the internal electric field is complex and nonuniform. Prototype tests of standard 3D sensors showed encouraging performance of $< 65$ps after noise subtraction\cite{5733383}. Field 
non-uniformity can be solved by replacing the round electrodes by 
trenches. Recent studies of such 3D trenched detectors measured a time resolution of 27ps\cite{Lai:2020dfg}.

The induced current from a track is essentially instantaneous. This provides a very fast rise current pulse whose magnitude and polarity depend on the weighting field for a given pixel. Pixel detectors with very low load and 
interconnect capacitance can utilize this initial current pulse to provide both fast timing and detailed information on the track location and angle\cite{Lipton:2019drv}. Extracting this information is a substantial challenge for the electronics, the pixel density is high, capacitance must be minimized, information must be extracted from a pixel field, and the patterns of information encoded in the current pulse are complex.

\subsection{Light Detectors}
Photodetectors have long been at the leading edge of fast timing. Cerenkov light and fast scintillators such as LYSO provide intrinsically fast signal sources. In the 20th century time resolution was typically determined by variations in electron travel time from the photocathode through the dynodes of photomultiplier tubes. The microchannel plate has reduced this time dispersion using a more compact geometry with micron-scale multiplication structures and has demonstrated ps-level timing \cite{10.1117/12.568180} in multiple applications. The LAPPD collaboration\cite{https://doi.org/10.48550/arxiv.1603.01843} reimagined the MCP, extending the area and lowering the cost of these devices by using ALD secondary emissive coating of extruded glass structures. The effort also involved development of novel photocathode formation and packaging techniques. An advantage of the LAPPD technology is the flexibility of the anode structures.  These can be pixelated, use charge sharing or charge division, or be capacitively coupled.  LAPPDs are now in limited production and have demonstrated time resolution less than of 50 ps\cite{Lyashenko:2019tdj}. LAPPDs are candidates for a number of applications across Particle and Nuclear Physics and are currently being tested in the ANNIE experiment at Fermilab\cite{Tiras:2019ozv}.

Semiconductor-based photodetectors such as SPADs and SIPMs can provide 10 ps-level time resolution for scintillation light\cite{Gundacker_2020} and charged particles. SIPMs are already widely used in HEP. The CMS Barrel Timing Layer (BTL)\cite{Butler:2019rpu} and High Granularity Calorimeter\cite{CMS:2017jpq} upgrades both incorporate SIPMs for fast timing. PET tomography studies include a roadmap to achieve 10 ps photon timing resolution\cite{Lecoq_2020}. SIPMs are constrained by the response speed, non-linearity and crosstalk inherent in their micro/marco pixel structure. Developments such as digital and 3D SIPMs\cite{8824571} can achieve improved capabilities, position, and time resolution by replacing the analog summing node with digital micropixels.

Other photodetector technologies under study which are capable of ps resolution include semiconductor quantum dot-based devices\cite{Oktyabrsky:2016ard}, ZbO:X ceramics \cite{https://doi.org/10.48550/arxiv.2203.09942} and others.

\subsection{Calorimeters}
Electromagnetic and hadron calorimetery present different requirements for detectors and electronics. Electromagnetic shower development is prompt and the shower shape is well-defined. By contrast hadronic showers take time to develop and have substantial energy and spatial fluctuations. A general description of timing-based approaches to these issues is discussed in the "Precision Timing for Collider Experiment based Calorimetry" whitepaper\cite{Chekanov:2022vyh}. Time resolution required depends on the application and range from sub-nanosecond to a few picoseconds.

Sensor systems for fast timing mentioned in the whitepaper including dual readout calorimeters, LGADs, CMOS MAPS, MCPs, Micromegas, LYSO, silicon or scintillator tiles, RPCs and coherent microwave Cerenkov detectors. Some of these, such as MAPS, integrate the sensing with the electronics. Others require conversion of signals, such as light to electronic pulses. Electronics challenges include:
\begin{itemize}
\item Very high system dynamic range (MIP to TeV?)
\item Achieving picosecond resolution in large, distributed multi-channel systems
\item System interconnection with preservation of time information
\item Time to Digital Conversion (TDC design)
\item Local processing and utilization of the signal and transmission of the results 
\item Waveform digitization and parameter extraction, particularly for dual readout calorimeters where much information is encoded in the waveform itself.
\end{itemize}
The various candidate sensor systems vary widely in scale, from $50\mu$m pixels in CMOS MAPS to cm-scale tiles in the CMS HGC, to larger volume dual readout and scintillator or crystal-based systems.

\section{ASIC R\&D}
%\subsection{Scaled CMOS}
There are several R\&D projects aimed at the development of fast ASIC electronics for future HEP applications.  We summarize the programs below and include technical details in appendices.
\begin{itemize}
    \item \textbf{SiGe amplifiers (A.1)} - Advances in detector technology and the direction of HEP experiments and applications require the development of new specialized readout electronics. A possible path to achieve O(10~ps) time resolution is an integrated chip using Silicon Germanium (SiGe) technology. Anadyne, Inc. in collaboration with University of California Santa Cruz has developed a prototype SiGe front end readout chip optimized for low power and timing resolution, with 0.5~mW per channel (front end and discriminator) while retaining 10~ps of timing resolution for 5~fC of injected charge.
    \item \textbf{Full Waveform digitization chip(A.2)} Time of arrival and time-over-threshold based readout strategies will likely adversely impact the ability to provide sub-pixel spatial resolution and typically have difficulty compensating for environmental factors such as pile-up, sensor aging, and radiation; timing precision can also be adversely impacted by factors such as timewalk, baseline wander and waveform shape variations. Full waveform digitization is expected to be more robust against a variety of adverse factors which can affect timing and spatial precision.
    \item \textbf{The FAST family of ASICs (A.3)} In the past several years the FAST effort had the goal of designing an ASICs tailored to the read-out of Ultra-Fast Silicon Detector.This family of ASICs aims to provide a 25~ps time resolution with rates up to 200 MHz and has been designed in a 110 nm CMOS commercial technology node. In every iteration of the production the architecture has been improved to optimize the chip performance. Starting from FAST2 a analog-only version of the chip have been produced.
    \item \textbf{ (A.4)} ETROC is the readout ASIC for the CMS Endcap Timing Layer, which has been under development at Fermilab with TSMC 65 nm CMOS technology. The development is divided into four prototyping phases, ETROC0, ETROC1, ETROC2 and ETROC3. ETROC0 consists of a single channel analog front-end with preamplifier and discriminator only, ETROC1 is a 4 × 4 array with full chain precision timing signal processing including a new time-to-digital convertor (TDC) for time of arrival (TOA) and time over threshold (TOT) measurements, while ETROC2 and ETROC3 are full size (16 × 16) and full functionality prototype chips.
    \item \textbf{ALTIROC family of ASICs} The ALTIROC ASIC, developed within the scope of the High-Granularity Timing Detector (HGTD) collaboration, has been developed for Atlas Phase II at the HL-LHC. ALTIROC1 has been designed on a CMOS 130 nm process and achieves a 25 ps time resolution, with 25 channels per ASIC. Each channel integrates an RF preamplifier, followed by a high speed discriminator and two TDCs for Time-of-Arrival and Time-Over-Threshold measurements, as well as a local memory~\cite{altiroc}. The TDC, designed by SLAC, is based on a Cycling Vernier Delay Line with a resolution of 20 ps and a maximum range of 2.5 ns (7 bits dynamic range).
    \item \textbf{CFD chip development at Fermilab} Given the importance of 4D tracking for future particle detectors, Fermilab is pursuing the development of a novel approach to time-stamping LGAD signals based on Constant Fraction Discriminator (CFD) approach. The aim is to develop a robust fast-timing discriminator for fast detectors, with a time resolution of 30ps or better, appropriate for use in IC pixels, stable and easy to use, and with very low dead-time ($\approx 25ns$)\cite{TZTWEPP}. 
    \item \textbf{28nm CMOS technology TDC design (A.6)} CERN has chosen the 28nm node based on radiation-hardness studies~\cite{Borghello2020}, frequency and cost of MPW runs and strong presence on the market. Furthermore, the 28nm technology is at least twice as fast and allows circuit densities around 4-5 times higher than the previously employed 65nm node, making it a good candidate for design of high granularity 4D trackers. SLAC has started the design of TDCs in 28nm technology node with target time resolutions of 10-50ps.
    \item \textbf{22nm CMOS technology TDC design (A.7)} Global Foundries 22 FDX has been used by Fermilab extensively for cryoelectronics development for both Quantum control and readout electronics as well as cryogenic detectors. Like the 28nm node, GF 22 FDX has small feature size and high $F_t$ and $F_{max}$, in addition to the uniquely high self gain and high current efficiency of fully-depleted silicon-on-insulator node (FDSOI), making this process an excellent candidate for highly integrated fast timing circuits.
    \item \textbf{3D Integration(A.8)} 3D integration of electronics (distinct from 3D sensor technology) is the interconnection of multiple layers of electronics and sensors using wafer thinning, hybrid bonding, and Through-Silicon-Vias (TSVs). This technology enables micron-level interconnections between layers of sensors and thinned electronics and deployment of heterogeneous stacks of sensors and readout.
    \item \textbf{Monolithic LGAD(A.9)}: the next logical step in the LGAD development  and in line with the trend in the silicon sensor industry is to integrate the sensitive volume with the read-out electronics. While modern CMOS technologies offer a wide range of possibilities and seem compatible with most of the LGAD process and front-end, some customizations appear unavoidable, for example in the addition of the gain layer. LGAD variants, such as AC-LGADs, should be compatible as well.
\end{itemize}

%\subsection{Ultrafast Sensor Readout Electronics}

%\subsection{TDC Design}
%\subsection{Power Constraints}
%\subsection{System Precision}

\section{R\&D Directions}
\subsection{Foundry Access}
The R\&D path for fast electronics must include development of foundry access for the appropriate nodes and technologies. CERN has been the leading lab over the last decade, qualifying the 65 nm node and negotiating access with TSMC through IMEC. Exploration of the 28 nm node has now begun and this will likely become the default choice over the next decade. Collaboration for fast electronics in this node along the RD53 model would be very beneficial to all parties. This provides a unified target for front end, TDC, and digital design collaboration, possible along the lines pioneered by RD53.

Although 28 nm bulk silicon may become the default process, many applications will require different process characteristics. Examples include SiGe for high speed, Fully Depleted Silicon On Insulator (FD-SOI) for speed and cryogenic operation, epitaxial and high voltage processes for MAPS, and legacy nodes for cost and simplicity. 

Especially valuable is the ability to work directly with foundries on process optimization for HEP applications. This capability is unusual, as foundry processes are closely held intellectual property. This access has been very valuable in developments such as HR-MAPS with Tower/Jazz, CCD development with MicroChip, and the 90 nm Skywater/MIT-LL process. It is important to maintain and develop these contacts.

\subsection{Front End}
The design and optimization of the front-end amplifier is particularly important for fast electronics.  The front end typically defines the signal/noise and thus the time jitter of the system. It can also consume significant power. Each design must be optimized for it's environment, considering signal source, input capacitance required resolution, and subsequent processing. 

Front end amplifiers in HEP are typically charge-sensitive. Full use for the input signal for fast detectors will likely require Trans-Impedance Amplifiers (TIA) to preserve the fast current signal. These require careful front-end design to maintain stability and low current draw. The design depends on the radiation and thermal environment and process parameters. Experienced analog designers will continue to be at a premium.

\subsection{Digitization}
Digitization of the amplified signal can take many forms depending on the signal source and application. For example signals from an standard LGAD are short with a uniform weighting field and may only require a time of arrival and time over threshold to define the pulse. Signals from a segmented detector or an AC LGAD may be more complex and may require multiple thresholds or waveform digitization to adequately characterize the input. 

There are a number of established TDC techniques including delay lines, multi-phase clock systems, vernier TDCs and time to amplitude conversion. These need to be adapted and extended for HEP applications and emerging processes. A central problem is system non-linearity caused by non-equal time bins over the sensitive interval. There are a number of linearization algorithms that can linearize TDC response and minimize overall power consumption.

TDC designs must continue to improve in precision and power consumption. Current TDC designs for ETL and HGTD serve mm$^2$-scale detectors with $\approx 200\mu$W
per channel.  Future designs might require TDC systems that deal with pixel sizes a factor of 20-40 smaller (400-1600x density). In this case a TDC per pixel may not be practical and changes in system architecture will be needed to provide fast timing information for a field of pixels. 

\subsection{Signal Processing}
The simultaneous need for position and time resolution in many applications adds an additional layer of complexity. For example a detector with intrinsic charge division, such as the AC-LGAD, will have both varying pulse heights and shapes as a function of position and distance from contacts (although rise times should be preserved). Pixelated detectors with thickness/pitch $>1$ will see varying pulse shapes as a function of electrode and position. For ultimate time resolution these effects must be compensated, either on-detector or as part of the processing chain. This is an opportunity to employ emerging technologies such as machine learning\cite{Guglielmo_2021} to front or back end systems to take advantage of all possible information.

\subsection{System Design}
Design of Particle Physics experiments are often dictated by the accelerator environment. For colliders, the crossing interval and luminosity often dictate the required time resolutions and processing and I/O budgets. This then defines the amount of information that can be sent off-detector. A highly granular detector may only be able to send Time-of-Arrival(TOA) and perhaps Time-over-Threshold(TOT) with the TOT capability dictated by crossing interval and occupancy. A longer interval between events per may allow more complex per pixel information, from multiple thresholds to waveform digitization. 

In experimental design the physics goals must be accomplished with a detector system that is buildable and affordable. Many applications are limited by power consumption at the front (input transistor current) and/or back ends (data transmission power). Fast timing applications put special emphasis on noise and rise time, which often requires higher front-end power. Increased pixel density to cope with required resolution and occupancy although somewhat balanced by lower load capacitance, also strains the power budget. Improved per hit time resolution may require more complex processing of the input waveform with corrections for delta rays, nonuniform ionization and varying weighting fields. This will require more complex, power hungry on-chip calculations or increased waveform information sent to downstream processing.

Most of the sub-75 ps systems demonstrated to date have either been in small, well constrained systems, or in beam tests\cite{Va_vra_2020}. Distribution and maintenance of the clock system will be a challenge for large systems. The CERN White Rabbit system design aims at " sub-nanosecond accuracy and picoseconds precision of synchronization for large distributed systems"\cite{WRabbit}. In large systems timing must be monitored and temperature and aging effects compensated. Optical transceivers can have several ps/degree delay variation. This has been considered in detail for Xilinx Ultrascale transievers and techniques have been developed to provide 1 ps phase resolution\cite{8967127}. Using a combination of these techniques and developments, including Constant Fraction Discrimination (CFD), waveform sampling and the latest generation of improved LAPPD, and combined with precise clock distribution allows for unprecedented accuracy in fast timing detectors. We anticipate that the grand challenge of measuring particles with a resolution of about 1ps is possible and would provide revolutionary physics opportunities.

%\subsection{Hadron Colliders}
%\subsection{Lepton Colliders}
%\subsection{EIC and Others}

\section{Conclusions}
Detectors and electronics are in a mutually supportive "arms race" moving toward ps resolution for HEP experiments. Rapidly evolving device scaling, 2.5 and 3D integration, digitization and TDC schemes, increased data bandwidth, new front and back end designs and innovative machine learning algorithms will all contribute. New fast signal sources such as LAPPDs, SIPMs and LGADs, induced current detectors, and micropattern gas detectors complement the electronics capabilities. 

Ultimate time resolution will require a detailed understanding of the experimental environment, optimization and matching of the signal source and front-end, compensation for variations in pulse shape, and utilisation of all available information in a pixelated sensor. Future experiments may require local filtering of complex events based on time and space information. Ultimately we may employ ASICs using sophisticated 3D architectures incorporating adaptive algorithms and machine learning techniques embedded in on-detector and triggering systems.

\appendix{}
\section{Appendix - Technical Descriptions}
\subsection{SiGe amplifiers}
Advances in detector technology and the direction of HEP experiments and applications require the development of new specialized readout electronics. 
Experimental demands include some combination of high rep rates (order of ns dead time), below 10 ps time of arrival (TOA) resolution, low power (between 0.1 mW and 1 mW per channel), and high dynamic range (for some specific application up to a few 1000s). 
A possible path to achieve O(10~ps) time resolution is an integrated chip using Silicon Germanium (SiGe) technology.
Using DoE SBIR funding, Anadyne, Inc. in collaboration with University of California Santa Cruz has developed a prototype SiGe front end readout chip optimized for low power and timing resolution, with 0.5~mW per channel (front end and discriminator) while retaining 10~ps of timing resolution for 5~fC of injected charge. 
In the process some insight was developed into the challenges and potential performance of SiGe front end ASICs for future R\&D effort. 
Channel matching to reduce calibration requirements and increase yield, timing resolution at the low end of the proposed detector dynamic range, and temperature stability were all considered during the design process to ensure the prototype performance would be deliverable in a full implementation. 
During this process we have developed some insight into the challenges and potential performance of SiGe front end ASICs if further R\&D were undertaken. 
The developed single pre-amplifier stage and what is effectively a Time Over Threshold (TOT) discriminator topology is suitable for low repetition rate and quiescent power and sub 10~ps timing resolution applications. 
The TOT data is required to correct the TOA of pulses over the entire dynamic range of interest. 
These TOT discriminators are not literal TOT converters of the amplified analog input. 
The output pulse width of the discriminator is proportional to the input pulse height and have a dead time of up to 10 ns. 
A constant fraction discriminator (CFD) may be more appropriate for applications with repetition rates greater than 100 MHz. 
One drawback of CFD schemes is that the pulse height information is lost, which can be useful for other purposes, such as determining interaction position in a segmented detector. 
An analysis of CFD dead times would be required to determine if they are in fact better for high rep rate applications. 
Some practical considerations for selecting a process for future R\&D include the size and power efficiency of the CMOS transistors for the back-end electronics and diminishing performance gains of higher speed SiGe transistors. 
The currently available SiGe processes offer 130~nm CMOS at a minimum. 
Transistors faster than 25~GHz have little signal to noise or power improvements to offer when designing readout systems for signals in the 1-2~GHz regime ultra-fast silicon detectors operate in. 
Moving to faster and smaller SiGe transistors may only introduce unnecessary design challenges such as poor transistor matching, low breakdown voltages, higher Vbe, etc. The current prototype is designed in a 10 GHz process. 
Significant R\&D efforts would be required to determine how much timing resolution, power consumption and dead time performance could be improved by moving to a specific 20-30 GHz process.

\subsection{Full digitization chip}
University of California Santa Cruz is currently working with Nalu Scientific, an ASIC design firm with experience developing readout solutions for HEP/NP, 
to design and fabricate a high channel density and scalable radiation-hard waveform digitization ASIC with embedded interface to advanced high-speed sensor arrays such as e.g. AC-LGADs. 
The chip is being fabricated with TSMC’s 65nm technology using design principles consistent with radiation hardening and targets the following features: picosecond-level timing resolution; 10~Gs/s waveform digitization rate 
to allow pulse shape discrimination; moderate data buffering (256~samples/chnl); 
autonomous chip triggering, readout control, calibration and storage virtualization; 
on-chip feature extraction and multi-channel data fusion; reduced cost and increased reliability due to embedded controller (reduction of external logic).
Existing readout approaches, such as ALTIROC~\cite{Angelo}  and the newer TimeSPOT1~\cite{Piccolo:2022hsz}, promise good-to-excellent timing resolution and channel density, and use a TDC-based measurement for signal arrival times and time-over-threshold (ToT) for an indirect estimate of integrated charge. 
However, these readout strategies will likely adversely impact the ability to provide sub-pixel spatial resolution and typically have difficulty compensating for environmental factors such as pile-up, sensor aging, and radiation; timing precision can also be adversely impacted by factors such as timewalk, baseline wander and waveform shape variations. 
Here, instead, full waveform digitization will be used, which is expected to be more robust against a variety of adverse factors which can affect timing and spatial precision. 

The initial iteration of the readout chip (v1) was recently (Jan 2022) fabricated for 50~um AC-LGADs. 
Later versions of the chip will be designed for 20~um pixel arrays and also test the minimum pitch feasible for a single-channel readout using a one-to-one pixel-input channel mapping.
Tests of v1 are planned over the next few months mainly to characterize the performance of (a) the input stage, with most channels implemented using a TIA but with one channel including TIA plus an internal amplifier, and (b) the full digitization chain, which is implemented in four different configurations with functional blocks that can be internally configured and connected in order to identify optimal digitization strategies. 
The final version of the chip will feature a transimpedance amplifier input stage able to be fine-tuned (or tunable) in order to accommodate high-density sensor arrays using technologies other than AC-LGADs. 
%Figure XYZ, below, shows the planned architecture for a mature sensor+chip package along with a blowup of the elements comprising a single readout channel, which will be tiled and mated to a sensor array to create the final fully integrated unit. 

\subsection{FAST family of ASICs}
In the past several years the FAST effort had the goal of designing an ASICs tailored to the read-out of Ultra-Fast Silicon Detector.
TOFFEE~\cite{TOFFEE}, the first prototype, has been produced in 2016, FAST1 in 2018\cite{FAST1}, and FAST2~\cite{FAST2} in 2020. 
This family of ASICs aims to provide a 25~ps time resolution with rates up to 200 MHz and has been designed in a 110 nm CMOS commercial technology node.
In every iteration of the production the architecture has been improved to optimize the chip performance. Starting from FAST2 a analog-only version of the chip have been produced.
All ASICs has been designed in a 110 nm CMOS commercial technology node and produced in multi-project wafer. 
All ASICs are optimized for an input capacitance of 3-6~pF, a range of temperature among -30 and +50 Celsius degrees and aim to provide a 25~ps time resolution with rates up to 200 MHz with a 6 pF UFSD.  
The next foreseen production is FAST3, which is based on the studies performed on FAST2 with expand linearity of the output dynamic range. 
FAST3\_Analog will have a redesigned output buffer to expand the linearity of the output dynamic range above 24~fC.
In parallel to FAST3, the ASIC UFSD\_ALCOR has been designed. It includes the optimized front-end stage used in FAST3\_Analog, a discriminator stage, time to digital converter (TDC), and a digital control unit.
The ASIC will be composed of 32 readout channels with a time resolution lower than 40 ps. Each channel operates at a maximum system clock frequency of 320~MHz. 
Each channel can measure the Time of Arrivals (ToA) and Time of Threshold (ToT) of a pulse signal with a least significant bit of 25~ps. 
The sample rate is around 1-2 MSa/s, depending on the configuration (ToA or ToT operation).
FAST3 and UFSD\_ALCOR are almost completed and will be manufactured in the first half of 2022. 
FAST1 and FAST2 are 1.7 mm × 5 mm chip consisting of 20 channels. 
In TOFFEE and FAST1 the channel architecture consists of a Trans-Impedance Amplifier (TIA), a second amplification stage based on a common source amplifier (CS), a two-stage leading edge discriminator (DISC1 and DISC2), a Pulse Width Regulator (PWR) to tune the digital output duration and a LVDS driver. 
FAST1 has been designed in three different flavors which differ in the front-end amplifier, REG, EVO1 and EVO2. 
The front-end amplifier used in the REG flavor, is based on a cascoded common source amplifier close in feedback by a resistor R of 20 k$\Omega$. 
This TIA stage is followed by a CS that provides a second amplification. The bandwidth of the front-end is 70 MHz and it allows increasing the noise-to-slope ratio by keeping the noise low. 
The EVO front-ends consist of a Broad-Band (BB) core amplifier close in feedback by a resistor which can be selected among 3 different values. A technology study has been included in the project, taping out EVO1 using standard transistors and EVO2 using RF transistors. 
Following the test of FAST1, the EVO1 and EVO2 architectures have been selected. The FAST2 production comprises three different ASICs: two (FAST2\_Digital\_EVO1, FAST2\_Digital\_EVO2) have 20 readout channels and implement an amplifier-comparator architecture, while the third ASIC (FAST2\_Analog) has 16 channels with only the amplification stage, 8 with the EVO1 front-end and 8 with the EVO2 front-end. 
FAST3, leveraging the studies performed on FAST2, will use the EVO1 technological choice. FAST3\_Analog will have a redesigned output buffer to expand the linearity of the output dynamic range above 24 fC.
In parallel to FAST3, the ASIC UFSD\_ALCOR has been designed. It includes the optimized front-end stage used in FAST3\_Analog, a discriminator stage, time to digital converter (TDC), and a digital control unit.  The ASIC will be composed of 32 readout channels with a time resolution lower than 40 ps. Each channel operates at a maximum system clock frequency of 320 MHz. Each channel can measure the Time of Arrivals (ToA) and Time of Threshold (ToT) of a pulse signal with a least significant bit of 25 ps. The sample rate is around 1-2 MSa/s, depending on the configuration (ToA or ToT operation). FAST3 and UFSD\_ALCOR are almost completed and will be manufactured in the first half of 2022. 

\subsection{ETROC family of ASICs}
The MIP Timing Detector (MTD) has been officially approved by the Compact Muon Solenoid (CMS) experiment for the High-Luminosity Large Hadron Collider (HL-LHC) upgrade. It is aiming to measure the arrival time of charged particles with a time resolution of 30 to 40 ps per track. The MTD consists of barrel and endcap sub-detectors. The endcap sub-detector or the endcap timing layer (ETL) is based on Low Gain Avalanche Detector (LGAD). It is designed to have two-layer hits for a given track such that the required time resolution per hit is in range between 40 to 50 ps. ETROC is the readout ASIC for ETL, which has been under development at Fermilab with TSMC 65 nm CMOS technology. The development is divided into four prototyping phases, ETROC0, ETROC1, ETROC2 and ETROC3. ETROC0 consists of a single channel analog front-end with preamplifier and discriminator only, ETROC1 is a 4 × 4 array with full chain precision timing signal processing including a new time-to-digital convertor (TDC) for time of arrival (TOA) and time over threshold (TOT) measurements, while ETROC2 and ETROC3 are the first full size (16 × 16) and full functionality prototype chip.

ETROC0 chips have been demonstrated in beam with time resolution of around 33 ps from the pre-amplifier waveform analysis and around 42 ps from the discriminator pulses analysis\cite{qsun2020etroc0}. A subset of ETROC0 chips have also been tested to a total ionizing dose (TID) of 100 MRad using X-ray machine at CERN and no performance degradation observed. ETROC0 design is successful, and the preamplifier and the discriminator are directly used in ETROC1 without modification. ETROC1 have a 4 × 4 pixel array with an H-tree style clock distribution network that is scalable to the final full size of 16 × 16. 

ETROC1 is the first full chain precision timing prototype, aiming to study and demonstrate the performance of the full signal chain, with the goal to achieve 40 to 50 ps time resolution per hit with LGAD (~30ps per track with two detector layer hits)\cite{qsun2021nss}. ETROC1 bare die has the dimension of 7 mm × 9mm. The signal chain in ETROC1 includes the preamplifier, the discriminator, the TDC and two readout paths, diagnostic readout and simple readout. In diagnostic readout, the TDC output of one of the 16 pixels is selected and readout at each bunch crossing clock period, while in simple readout mode, the TDC output in each pixel is stored in an in-pixel memory and the selected pixels are readout when an L1 acceptance signal is received.
One of the challenges of the ETROC design is that the TDC is required to consume less than 200 µW for each pixel at the nominal hit occupancy of 1\%. To meet the low-power requirement, the ETROC team uses a single delay line for both the TOA and the TOT measurements without delay control\cite{wzhang2021tdc}. This TDC is based on a simple delay-line approach originally developed in FPGA implementation. A double-strobe self-calibration scheme is used to calibrate TDC bin size under process, temperature, and power supply voltage variation. The overall performances of the TDC have been evaluated and meet the CMS ETL upgrade requirements. The TOA has a bin size of 17.8 ps within its effective dynamic range of 11.6 ns. The TOT has a bin size of 35.4 ps within its measured dynamic range of 9.8 ns. The effective measurement precisions of the TDC are 5.6 ps and 9.9 ps for the TOA and 10.4 ps and 16.7 ps for the TOT with and without the nonlinearity correction, respectively.

The bare ETROC1 chips have been tested extensively using charge injection, and the measured performance agrees well with the expectation, including the power consumptions. Some of bump bonded ETROC1 chips (with LGAD sensors) have been also extensively tested using charge injection, laser and test beam, respectively. The timing performance with the full signal chain as well as the 4 × 4 pixel array clock distribution network has been studied. Less than 40 ps time resolution was obtained with laser input. A three-board telescope with ETROC1 and LGAD was built and tested in Fermilab Test Beam Facility. The time resolution of ETROC1+LGAD from off-line data analysis is between 42 ps and 46 ps.

ETROC2 is planned to be submitted in summer 2022, which includes the 16 × 16 pixel array with full-chain readout and supporting blocks, e.g. I2C, PLL, Efuse, Temperature sensor, reference voltage generator, fast command decoder, and etc. Each pixel has an in-pixel threshold calibration\cite{Sun:2021rlz} block which helps calibrate discriminator threshold voltage. A circular buffer matching L1 latency is included in each pixel. A scalable switching network is developed to readout data and trigger from TDC. Two serial links, each up to 1.28 GHz, are used to send data and trigger out.
The next phase of ETROC prototyping will be ETROC3. The production of ETROC is foreseen to happen after 2024.

\subsection{CFD chip development at Fermilab}

Given the importance of 4D tracking for future particle detectors, Fermilab is pursuing the development of a novel approach to time-stamping LGAD signals based on Constant Fraction Discriminator (CFD) approach. 
The aim is to develop a robust fast-timing discriminator for fast detectors, with a time resolution of 30pS or better, appropriate for use in IC pixels, stable and easy to use, and with very low dead-time (~ 25ns). 
The Fermilab CFD v0 chip (FCFD0), designed in 65nm CMOS, uses several new techniques to achieve low power, area, jitter, time walk, and drift. This enables a simple and robust timing measurement ($\sim$30ps) of LGAD signals that vary in amplitude by at least a factor of 10, with no critical threshold setting or corrections required.

Precise measurements and calibrations of the chip on a bench, have confirmed stable operations, low dead time, consistent with simulations ($\sim$30ps at 5fC, and $<$ 10ps at 30fC).

\subsection{28nm CMOS technology TDC design}
CERN’s EP R\&D WP5: CMOS Technologies~\cite{CERN-EP-RDET-2021-001} survey has promoted the selection of 28nm CMOS node as the next step in microelectronics scaling for HEP designs. The choice was based on radiation-hardness studies~\cite{Borghello2020}, frequency and cost of MPW runs and strong presence on the market. Furthermore, the 28nm technology is at least twice as fast and allows circuit densities around 4-5 times higher than the previously employed 65nm node, making it a good candidate for design of high granularity 4D trackers. One of the critical circuit blocks necessary to enable 4D operation in trackers are low-power and compact Time-to-Digital Converters (TDC) capable of high time-measurement precision. SLAC has stated the design of TDCs in 28nm technology node with target time resolutions of 10-50ps. The plan is to submit the first prototype for fabrication at the end of this year.

\subsection{22nm CMOS technology TDC design}
GF 22 FDX has been used by Fermilab extensively for cryoelectronics development for both Quantum control and readout electronics as well as cryogenic detectors. 

Just like the 28nm node, GF 22 FDX has small feature size and high $F_t$ and $F_{max}$, in addition to the uniquely high self gain and high current efficiency of fully-depleted silicon-on-insulator node (FDSOI), making this process an excellent candidate for highly integrated fast timing circuits. 
The availability of a backgate can be used to improve the performance of the devices, for example by negating the increase in threshold voltage in cryogenic environments; while the inherent radiation hardness, although not as extreme as for the 28nm bulk node, is aided by the extremely thin insulator.

Fermilab is currently developing multi-channel, low-jitter TDC ASICs for the readout of Superconducting Nanowires Single Photon Detectors (SNSPD). 
While this class of detectors, which has demonstrated sub-3ps jitter, has found widespread application in quantum imaging and photon counting, Fermilab is leading a large collaborative co-design effort to develop particle detectors based on this technology, with the aim of developing 4D detectors by leveraging the excellent timing properties of the nanowires.
In Decembre 2021, we prototyped the first version of a single channel TDC for cryogenic operation at 4K, targeting 5ps resolution, $>$10ns dynamic range, and $<0.5mW$ of power. Delivery is expected by April 2022.

\subsection{3D Integration}
3D integration of electronics (distinct from 3D sensor technology) is the interconnection of multiple layers of electronics and sensors using wafer thinning, hybrid bonding, and Through-Silicon-Vias (TSVs). This technology enables micron-level interconnections between layers of sensors and thinned electronics and deployment of heterogeneous stacks of sensors and readout. The first demonstration of this technology, now widely utilized for imaging, in HEP was through a FNAL-led collaboration, which produced a 3-tier stack of a sensor and two ROIC layers. This demonstrated fine pitch interconnect ($3\mu$m),  substantially lower interconnect capacitance than bump bonding, separation of analog and digital tiers, and edge-less readout\cite{Lipton:2015vca}.

For fast timing applications, the low input capacitance and fine pitch enabled by these technologies can provide high initial signal/noise as well as low front-end power. The combination of low input capacitance, a dedicated analog layer, and digital processing and I/O tiers can enable sophisticated processing of fields of pixels enabling on-detector fast timing and pattern recognition. 

\subsection{Monolithic LGADs}
As already happened with pixel detectors, also in the case of LGAD it is convenient to integrate sensing and readout elements onto the same substrate and have it fabricated by commercial CMOS foundries. These new structures will also avoid the need of interconnecting the sensor to the read-out electronics, a step that can be time-consuming, expensive, and prone to errors. CMOS foundries, on the other hand, deliver state of the art devices, cost-optimized, and with excellent yield. The CMOS technology must feature a high resistivity substrate, a few tens of microns thick and customized at the very least to allocate a gain p-type layer under the deepest n-well, while circuitry can stay in the inner p and n-wells as usual. Besides LGADs, also AC-LGADs and Deep-Junction LGADs can be fabricated in CMOS, provided the needed degree of customization is permitted into the standard production line. BNL is currently exploring if this effort is viable, by looking for suitable CMOS technologies whose process is compatible with monolithic LGADs and, at the same time,  allow the needed customization.

\bibliographystyle{unsrt}
\bibliography{bibliography.bib}

%%%%%%%%%%%%%%%%%%%%%%%%%%%%%%%%%%%%%%%%%%%%%%%%%%%%%%%%%%%%%%%%%%%%%%%%%
% example figure

%\begin{figure}
%\begin{center}
%\includegraphics[width=0.40\hsize]{xxx}
%\end{center}
%\caption{xxx}
%\label{fig:xxx}
%\end{figure}

%%%%%%%%%%%%%%%%%%%%%%%%%%%%%%%%%%%%%%%%%%%%%%%%%%%%%%%%%%%%%%%%%%%%%%%%%%%

%%%%%%%%%%%%%%%%%%%%%%%%%%%%%%%%%%%%%%%%%%

%  If you would like to use BibTEX for the bibliography, please feel free to do so.  It is not required.

%  To use BibTeX,

%    1.  uncomment the following two lines, 
%    2.  comment out everything below from  \begin{thebibliography}{99}   to \end{thebibliography).
%    3.  create the file  myreferences.bib, and process this file in the usual way

%\bibliographystyle{JHEP}
%\bibliography{myreferences}  % file myreferences.bib

%%%%%%%%%%%%%%%%%%%%%%%%%%%%%%%%%%%%%%%%%
%example bibliography

%\begin{thebibliography}{99}

%  this is a vanilla LaTeX bibliography.  It is also fine to use
%  bibTeX, but please be sure that bibTeX does not mangle your
%  citations

%\bibitem{}

%\end{thebibliography}

\end{document}

%% file: workshopsymbols.tex
\def\beq{\begin{equation}}
\def\eeq#1{\label{#1}\end{equation}}
\def\eeqn{\end{equation}}

\newenvironment{Eqnarray}%
   {\arraycolsep 0.14em\begin{eqnarray}}{\end{eqnarray}}
\def\beqa{\begin{Eqnarray}}
\def\eeqa#1{\label{#1}\end{Eqnarray}}
\def\eeqan{\end{Eqnarray}}

\let\bar=\overbar

\def\lsim{\mathrel{\raise.3ex\hbox{$<$\kern-.75em\lower1ex\hbox{$\sim$}}}}
\def\gsim{\mathrel{\raise.3ex\hbox{$>$\kern-.75em\lower1ex\hbox{$\sim$}}}}

\def\del{\partial}
\def\Dslash{\not{\hbox{\kern-4pt $D$}}}
\def\dslash{\not{\hbox{\kern-2pt $\del$}}}
\def\pslash{\not{\hbox{\kern-2pt $p$}}}
\def\ETmiss{\not{\hbox{\kern-4pt $E$}}_T}

\def\Dlr{\mathrel{\raise1.5ex\hbox{$\leftrightarrow$\kern-1em\lower1.5ex\hbox{$D$}}}}

\def\MSB{{\bar{M \kern -2pt S}}}
\def\msb{{\bar{\scriptsize M \kern -1pt S}}}

\def\drb{{\bar{\scriptsize D \kern -1pt R}}}

%% file: IF07_topic4.bbl
\begin{thebibliography}{10}

\bibitem{https://doi.org/10.48550/arxiv.2203.07224}
Sergo Jindariani, Federico Meloni, Nadia Pastrone, Chiara Aimè, Nazar
  Bartosik, Emanuela Barzi, Alessandro Bertolin, Alessandro Braghieri, Laura
  Buonincontri, Simone Calzaferri, Massimo Casarsa, Maria~Gabriella Catanesi,
  Alessandro Cerri, Grigorios Chachamis, Anna Colaleo, Camilla Curatolo,
  Giacomo Da~Molin, Jean-Pierre Delahaye, Biagio Di~Micco, Tommaso Dorigo,
  Filippo Errico, Davide Fiorina, Alessio Gianelle, Carlo Giraldin, John
  Hauptman, Tova~Ray Holmes, Karol Krizka, Lawrence Lee, Kenneth Long,
  Donatella Lucchesi, Nikolai Mokhov, Alessandro Montella, Federico Nardi, Niko
  Neufeld, David Neuffer, Simone~Pagan Griso, Antonello Pellecchia, Karolos
  Potamianos, Emilio Radicioni, Raffaella Radogna, Cristina Riccardi, Luciano
  Ristori, Lucio Rossi, Paola Salvini, Daniel Schulte, Lorenzo Sestini,
  Vladimir Shiltsev, Federica~Maria Simone, Anna Stamerra, Xiaohu Sun, Jian
  Tang, Emily~Anne Thompson, Ilaria Vai, Nicolo' Valle, Rosamaria Venditti,
  Piet Verwilligen, Paolo Vitulo, Hannsjorg Weber, Katsuya Yonehara, Angela
  Zaza, and Davide Zuliani.
\newblock Promising technologies and r\&d directions for the future muon
  collider detectors, 2022.

\bibitem{Cartiglia:2016sjr}
N.~Cartiglia et~al.
\newblock {The 4D pixel challenge}.
\newblock {\em JINST}, 11(12):C12016, 2016.

\bibitem{Spieler:1982jf}
H.~Spieler.
\newblock {FAST TIMING METHODS FOR SEMICONDUCTOR DETECTORS. (TALK)}.
\newblock 6 1982.

\bibitem{Sadrozinski:2017qpv}
Hartmut F.~W. Sadrozinski, Abraham Seiden, and Nicol\`o Cartiglia.
\newblock {4D tracking with ultra-fast silicon detectors}.
\newblock {\em Rept. Prog. Phys.}, 81(2):026101, 2018.

\bibitem{https://doi.org/10.48550/arxiv.2201.07772}
Ryan Heller, Christopher Madrid, Artur Apresyan, William~K. Brooks, Wei Chen,
  Gabriele D'Amen, Gabriele Giacomini, Ikumi Goya, Kazuhiko Hara, Sayuka Kita,
  Sergey Los, Adam Molnar, Koji Nakamura, Cristián Peña, Claudio~San Martín,
  Alessandro Tricoli, Tatsuki Ueda, and Si~Xie.
\newblock Characterization of bnl and hpk ac-lgad sensors with a 120 gev proton
  beam, 2022.

\bibitem{RD51}
Rd51.
\newblock https://rd51-public.web.cern.ch/.

\bibitem{BORTFELDT2018317}
J.~Bortfeldt, F.~Brunbauer, C.~David, D.~Desforge, G.~Fanourakis, J.~Franchi,
  M.~Gallinaro, I.~Giomataris, D.~González-Díaz, T.~Gustavsson, C.~Guyot,
  F.J. Iguaz, M.~Kebbiri, P.~Legou, J.~Liu, M.~Lupberger, O.~Maillard,
  I.~Manthos, H.~Müller, V.~Niaouris, E.~Oliveri, T.~Papaevangelou,
  K.~Paraschou, M.~Pomorski, B.~Qi, F.~Resnati, L.~Ropelewski, D.~Sampsonidis,
  T.~Schneider, P.~Schwemling, L.~Sohl, M.~van Stenis, P.~Thuiner,
  Y.~Tsipolitis, S.E. Tzamarias, R.~Veenhof, X.~Wang, S.~White, Z.~Zhang, and
  Y.~Zhou.
\newblock Picosec: Charged particle timing at sub-25 picosecond precision with
  a micromegas based detector.
\newblock {\em Nuclear Instruments and Methods in Physics Research Section A:
  Accelerators, Spectrometers, Detectors and Associated Equipment},
  903:317--325, 2018.

\bibitem{6154414}
S.~Martoiu, H.~Muller, and J.~Toledo.
\newblock Front-end electronics for the scalable readout system of rd51.
\newblock In {\em 2011 IEEE Nuclear Science Symposium Conference Record}, pages
  2036--2038, 2011.

\bibitem{Barbosa2022Snowmass2I}
F.~T.~F. Barbosa, Daniel Bazin, Francesco Boss'u, M.~Cortesi, Silvia~Dalla
  Torre, Sergey Furletov, Y.~Furletova, Paul Gu{\`e}ye, Kondo Gnanvo, Marcus
  Hohlmann, Wolfgang Mittig, Damien Neyret, Matthiew Posik, and C.~Wrede.
\newblock Snowmass 2021 instrumentation frontier (if5 - mpgds) -- white paper
  2: Micro pattern gaseous detectors for nuclear physics.
\newblock 2022.

\bibitem{https://doi.org/10.48550/arxiv.2203.06562}
K.~Dehmelt, M.~Della~Pietra, H.~Muller, S.~E. Tzamarias, A.~White, S.~White,
  Z.~Zhang, M.~Alviggi, I.~Angelis, S.~Aune, J.~Bortfeldt, M.~Bregant,
  F.~Brunbauer, M.~T. Camerlingo, V.~Canale, V.~D'Amico, D.~Desforge,
  C.~Di~Donato, R.~Di~Nardo, G.~Fanourakis, K.~J. Floethner, M.~Gallinaro,
  F.~Garcia, I.~Giomataris, K.~Gnanvo, T.~Gustavsson, R.~Hall-Wilton, P.~Iengo,
  F.~J. Iguaz, M.~Iodice, D.~Janssens, A.~Kallitsopoulou, M.~Kebbiri,
  K.~Kordas, C.~Lampoudis, P.~Legou, M.~Lisowska, J.~Liu, M.~Lupberger,
  S.~Malace, I.~Maniatis, I.~Manthos, Y.~Meng, H.~Natal da~Luz, E.~Oliveri,
  G.~Orlandini, T.~Papaevangelou, K.~Paraschou, F.~Petrucci, D.~Pfeiffer,
  M.~Pomorski, S.~Popescu, F.~Resnati, L.~Ropelewski, A.~Rusu, D.~Sampsonidis,
  L.~Scharenberg, T.~Schneider, G.~Sekhniaidze, M.~Sessa, M.~Shao, L.~Sohl,
  J.~Toledo-Alarcon, A.~Tsiamis, Y.~Tsipolitis, A.~Utrobicic, M.~van Stenis,
  R.~Veenhof, X.~Wang, and Y.~Zhou.
\newblock Snowmass 2021 white paper instrumentation frontier 05 -- white paper
  1: Mpgds: Recent advances and current r\&d, 2022.

\bibitem{AglieriRinella:2019eri}
G.~Aglieri~Rinella et~al.
\newblock {The NA62 GigaTracKer: a low mass high intensity beam 4D tracker with
  65 ps time resolution on tracks}.
\newblock {\em JINST}, 14:P07010, 2019.

\bibitem{5733383}
Sherwood Parker, Angela Kok, Christopher Kenney, Pierre Jarron, Jasmine Hasi,
  Matthieu Despeisse, Cinzia Da~Vià, and Giovanni Anelli.
\newblock Increased speed: 3d silicon sensors; fast current amplifiers.
\newblock {\em IEEE Transactions on Nuclear Science}, 58(2):404--417, 2011.

\bibitem{Lai:2020dfg}
A.~Lai.
\newblock {Timing characterisation of 3D-trench silicon sensors}.
\newblock {\em JINST}, 15(09):C09054, 2020.

\bibitem{Lipton:2019drv}
Ronald Lipton and Jason Theiman.
\newblock {Fast timing with induced current detectors}.
\newblock {\em Nucl. Instrum. Meth. A}, 945:162423, 2019.

\bibitem{10.1117/12.568180}
James~S. Milnes and J.~Howorth.
\newblock {Picosecond time response characteristics of microchannel plate PMT
  detectors}.
\newblock In Dennis~L. Paisley, Stuart Kleinfelder, Donald~R. Snyder, and
  Brian~J. Thompson, editors, {\em 26th International Congress on High-Speed
  Photography and Photonics}, volume 5580, pages 730 -- 740. International
  Society for Optics and Photonics, SPIE, 2005.

\bibitem{https://doi.org/10.48550/arxiv.1603.01843}
Bernhard~W. Adams, Klaus Attenkofer, Mircea Bogdan, Karen Byrum, Andrey Elagin,
  Jeffrey~W. Elam, Henry~J. Frisch, Jean-Francois Genat, Herve Grabas, Joseph
  Gregar, Elaine Hahn, Mary Heintz, Zinetula Insepov, Valentin Ivanov, Sharon
  Jelinsky, Slade Jokely, Sun~Wu Lee, Anil.~U. Mane, Jason McPhate, Michael~J.
  Minot, Pavel Murat, Kurtis Nishimura, Richard Northrop, Razib Obaid, Eric
  Oberla, Erik Ramberg, Anatoly Ronzhin, Oswald~H. Siegmund, Gregory Sellberg,
  Neal~T. Sullivan, Anton Tremsin, Gary Varner, Igor Veryovkin, Alexei
  Vostrikov, Robert~G. Wagner, Dean Walters, Hsien-Hau Wang, Matthew Wetstein,
  Junqi Xi, Zikri Yusov, and Alexander Zinovev.
\newblock A brief technical history of the large-area picosecond photodetector
  (lappd) collaboration, 2016.

\bibitem{Lyashenko:2019tdj}
A.~V. Lyashenko et~al.
\newblock {Performance of Large Area Picosecond Photo-Detectors
  (LAPPD$^{TM}$)}.
\newblock {\em Nucl. Instrum. Meth. A}, 958:162834, 2020.

\bibitem{Tiras:2019ozv}
Emrah Tiras.
\newblock {Detector R\&D for ANNIE and Future Neutrino Experiments}.
\newblock In {\em {Meeting of the Division of Particles and Fields of the
  American Physical Society}}, 10 2019.

\bibitem{Gundacker_2020}
Stefan Gundacker and Arjan Heering.
\newblock The silicon photomultiplier: fundamentals and applications of a
  modern solid-state photon detector.
\newblock {\em Physics in Medicine {\&} Biology}, 65(17):17TR01, aug 2020.

\bibitem{Butler:2019rpu}
Joel~N. Butler and Tommaso Tabarelli~de Fatis.
\newblock {A MIP Timing Detector for the CMS Phase-2 Upgrade}.
\newblock 2019.

\bibitem{CMS:2017jpq}
{The Phase-2 Upgrade of the CMS Endcap Calorimeter}.
\newblock 2017.

\bibitem{Lecoq_2020}
Paul Lecoq, Christian Morel, John~O Prior, Dimitris Visvikis, Stefan Gundacker,
  Etiennette Auffray, Peter Kri{\v{z}}an, Rosana~Martinez Turtos, Dominique
  Thers, Edoardo Charbon, Joao Varela, Christophe de~La~Taille, Angelo Rivetti,
  Dominique Breton, Jean-Fran{\c{c}}ois Pratte, Johan Nuyts, Suleman Surti,
  Stefaan Vandenberghe, Paul Marsden, Katia Parodi, Jose~Maria Benlloch, and
  Mathieu Benoit.
\newblock Roadmap toward the 10 ps time-of-flight {PET} challenge.
\newblock {\em Physics in Medicine {\&} Biology}, 65(21):21RM01, oct 2020.

\bibitem{8824571}
Samuel Parent, Maxime Côté, Frédéric Vachon, Robert Groulx, Stéphane
  Martel, Henri Dautet, Serge~A. Charlebois, and Jean-François Pratte.
\newblock Single photon avalanche diodes and vertical integration process for a
  3d digital sipm using industrial semiconductor technologies.
\newblock In {\em 2018 IEEE Nuclear Science Symposium and Medical Imaging
  Conference Proceedings (NSS/MIC)}, pages 1--4, 2018.

\bibitem{Oktyabrsky:2016ard}
Serge Oktyabrsky, Michael Yakimov, Vadim Tokranov, and Pavel Murat.
\newblock {Integrated Semiconductor Quantum Dot Scintillation Detector:
  Ultimate Limit for Speed and Light Yield}.
\newblock {\em IEEE Trans. Nucl. Sci.}, 63(2):656--663, 2016.

\bibitem{https://doi.org/10.48550/arxiv.2203.09942}
David~R Winn, Y.~Onel, and B.~Bilki.
\newblock High rate and high precision timing and calorimeter detectors, 2022.

\bibitem{Chekanov:2022vyh}
S.~V. Chekanov et~al.
\newblock {Precision timing for collider-experiment-based calorimetry}.
\newblock 3 2022.

\bibitem{altiroc}
C.~Agapopoulou, P.~Dinaucourt, A.~Dragone, D.~Gong, C.~de~La~Taille,
  N.~Makovec, B.~Markovic, G.~Martin-Chassard, C.~Milke, M.~Morenas,
  L.~Ruckman, S.~Sacerdoti, A.~Schwartzman, N.~Seguin-Moreau, L.~Serin, D.~Su,
  and J.~Ye.
\newblock Altiroc 1, a 25 ps time resolution asic for the atlas high
  granularity timing detector.
\newblock In {\em 2020 IEEE Nuclear Science Symposium and Medical Imaging
  Conference (NSS/MIC)}, pages 1--4, 2020.

\bibitem{TZTWEPP}
{T. Zimmerman}.
\newblock Precision timing asic for lgad sensors based on a constant fraction
  discriminator – fcfd0, 2021.
\newblock
  \url{https://indico.cern.ch/event/1019078/contributions/4443948/attachments/2277824/3938152/FCFD0_TWEPP_talk.pdf},
  Last accessed on 2022-03-29.

\bibitem{Borghello2020}
{Giulio Borghello, “Ionizing Radiation Effects On 28 nm CMOS Technology”}.
\newblock Technical report, CERN, Geneva, May 2020.

\bibitem{Guglielmo_2021}
Giuseppe~Di Guglielmo, Farah Fahim, Christian Herwig, Manuel~Blanco Valentin,
  Javier Duarte, Cristian Gingu, Philip Harris, James Hirschauer, Martin Kwok,
  Vladimir Loncar, Yingyi Luo, Llovizna Miranda, Jennifer Ngadiuba, Daniel
  Noonan, Seda Ogrenci-Memik, Maurizio Pierini, Sioni Summers, and Nhan Tran.
\newblock A reconfigurable neural network {ASIC} for detector front-end data
  compression at the {HL}-{LHC}.
\newblock {\em {IEEE} Transactions on Nuclear Science}, 68(8):2179--2186, aug
  2021.

\bibitem{Va_vra_2020}
J~Va'vra.
\newblock Picosecond timing detectors and applications.
\newblock {\em Journal of Physics: Conference Series}, 1498(1):012013, apr
  2020.

\bibitem{WRabbit}
"J. Serrano and others".
\newblock The white rabbit project.
\newblock In {\em Proceedings of ICALEPCS}, 2009.

\bibitem{8967127}
Eduardo Mendes, Sophie Baron, Csaba Soos, Jan Troska, and Paolo Novellini.
\newblock Achieving picosecond-level phase stability in timing distribution
  systems with xilinx ultrascale transceivers.
\newblock {\em IEEE Transactions on Nuclear Science}, 67(3):473--481, 2020.

\bibitem{Angelo}
B.~Markovic et~al.
\newblock {ALTIROC1}, a 20 ps time-resolution asic prototype for the atlas high
  granularity timing detector {HGTD}.
\newblock {\em 2018 IEEE Nuclear Science Symposium and Medical Imaging
  Conference Proceedings (NSS/MIC), Sydney, Australia}, page 1–3, Dec 2018.

\bibitem{Piccolo:2022hsz}
Lorenzo Piccolo, Sandro Cadeddu, Luca Frontini, Adriano Lai, Valentino
  Liberali, Angelo Rivetti, and Alberto Stabile.
\newblock {First Measurements on the Timespot1 ASIC: a Fast-Timing, High-Rate
  Pixel-Matrix Front-End}.
\newblock 1 2022.

\bibitem{TOFFEE}
F.~Cenna, N.~Cartiglia, A.~Di Francesco, J.~Olave, M.~Da~Rocha Rolo,
  A.~Rivetti, J.C. Silva, R.~Silva, and J.~Varela.
\newblock {TOFFEE}: a full custom amplifier-comparator chip for timing
  applications with silicon detectors.
\newblock {\em Journal of Instrumentation}, 12(03):C03031--C03031, mar 2017.

\bibitem{FAST1}
E.J. Olave, F.~Fausti, N.~Cartiglia, R.~Arcidiacono, H.F.-W. Sadrozinski, and
  A.~Seiden.
\newblock Design and characterization of the fast chip: a front-end for 4d
  tracking systems based on ultra-fast silicon detectors aiming at 30 ps time
  resolution.
\newblock {\em Nuclear Instruments and Methods in Physics Research Section A:
  Accelerators, Spectrometers, Detectors and Associated Equipment}, 985:164615,
  2021.

\bibitem{FAST2}
{A. Rojas, FAST2: a new family of front-end ASICs to read out thin Ultra-Fast
  Silicon detectors achieving picosecond time resolution.}
\newblock {\url{https://indico.cern.ch/event/1019078/contributions/4443951/}}.

\bibitem{qsun2020etroc0}
Quan Sun et~al.
\newblock The analog front-end for the lgad based precision timing application
  in cms etl, 2020.

\bibitem{qsun2021nss}
Quan Sun et~al.
\newblock Etroc1: the first full chain precision timing prototype for cms mtd
  endcap timing layer.

\bibitem{wzhang2021tdc}
Wei Zhang et~al.
\newblock A low-power time-to-digital converter for the cms endcap timing layer
  (etl) upgrade.
\newblock {\em IEEE Transactions on Nuclear Science}, 68(8):1984--1992, 2021.

\bibitem{Sun:2021rlz}
Hanhan Sun et~al.
\newblock {In-pixel automatic threshold calibration for the CMS Endcap Timing
  Layer readout chip}.
\newblock {\em JINST}, 16(09):T09006, 2021.

\bibitem{CERN-EP-RDET-2021-001}
{Strategic R\&D Programme on Technologies for Future Experiments - Annual
  Report 2020}.
\newblock Technical Report CERN-EP-RDET-2021-001, CERN, Geneva, Apr 2021.

\bibitem{Lipton:2015vca}
Ronald Lipton et~al.
\newblock {Three Dimensional Integrated Circuits Bonded to Sensors}.
\newblock {\em PoS}, Vertex2014:045, 2015.

\end{thebibliography}
